\pdfoutput=1
\documentclass[12 pt]{article}
\usepackage[english,ukrainian,russian]{babel}
\usepackage[utf8]{inputenc}

\usepackage[a4paper,top=3cm,bottom=2cm,left=3cm,right=3cm,marginparwidth=1.75cm]{geometry}

\usepackage{graphicx}

\begin{document}

\title{THIRD COMPONENTS WITH ELLIPTICAL ORBITS IN THE ECLIPSING BINARIES}
\author{\textsl{D.E. Tvardovskyi$^{1, 2}$, V.I. Marsakova$^{1}$, I.L. Andronov$^2$}}
\date{\vspace*{-6ex}}
\maketitle
$^1$ Odessa I.I. Mechnikov National University, Odessa, Ukraine\\
$^2$ Department ''Mathematics, Physics and Astronomy'', Odessa National Maritime University Odessa, Ukraine

\begin{abstract}
In our research, we studied 9 eclipsing binary stars: AR Lac, U CrB, S Equ, SU Boo, VV UMa, WW Gem, YY Eri, V0404 Lyr, HP Aur. We collected large sets of moments of minima from BRNO and observational data from AAVSO databases. Then, we obtained moments of minima for AAVSO observations (totally - 397 minima) using method of approximation with symmetric polynomial, realized in software MAVKA. This software was provided by Kateryna D. Andrych and Ivan L. Andronov. After we combined obtained moments of minima with data taken from BRNO and plotted O-C diagrams. For all stars these diagrams represented sinusoidal-like oscillations with superposition of linear (for SU Boo, VV UMa, WW Gem, V0404 Lyr and HP Aur) or parabolic trend (for AR Lac, U CrB, S Equ and YY Eri). The oscillations could be described as presence of the third component, which does not take part in eclipses, but causes the well-known light-time effect (LTE). That effect could be easily detected using long data series of observations. However, the oscillations have clear asymmetry, which we interpreted as elliptical shape of the third components orbit. Parabolic trend we explained as mass transfer between components of binary system. For all stars we computed minimal possible mass of the third component. In addition, we developed our own code in computed language Python and using it we computed orbital elements of the third component. Moreover, for stars with parabolic trend we calculated rate of the mass transfer. Finally, for all computed values we estimated errors.
\\[3mm]
В нашому дослідженні ми вивчили 9 затемнюваних подвійних зорь: AR Lac, U CrB, S Equ, SU Boo, VV UMa, WW Gem, YY Eri, V0404 Lyr, HP Aur. Ми зібрали великі масиви мінімумів із BRNO та спостережень із бази даних AAVSO. Потім ми отримали моменти мінімумів для спостережень AAVSO (всього - 397 мінімумів) за допомогою методу апроксимації з симетричним поліномом, реалізованого в програмі MAVKA. Це програмне забезпечення надали Катерина Д. Андрич та Іван Л. Андронов. Після того ми поєднали отримані моменти мінімумів із даними, отриманими з BRNO, та побудували діаграми O-C. Для всіх зорь ці діаграми представляли собою подібні до синусоїдальних коливання з суперпозицією лінійного (для SU Boo, VV UMa, WW Gem, V0404 Lyr та HP Aur) або параболічного тренду (для AR Lac, U CrB, S Equ та YY Eri). Коливання можна описати як наявність третього компонента, який не бере участі в затемненнях, але викликає відомий light-time effect (LTE). Цей ефект можна легко виявити, використовуючи довгі серії даних спостережень. Однак коливання мають чітку асиметрію, яку ми інтерпретували як еліптичну форму орбіти третього компонента. Параболічний тренд ми пояснили, як перетікання речовини між компонентами подвійної системи. Для всіх зорь ми обчислили мінімально можливу масу третього компонента. Крім того, ми розробили власний код на мові програмування Python та за його допомогою ми обчислили елементи орбіти третього компонента. Більше того, для зорь з параболічним трендом ми розраховували швидкість перетікання речовини. В решті-решт, для всіх обчислених значень ми оцінили похибки.
\end{abstract}
\section{Introduction}
\subsection{General information}
For this research we chose 9 eclipsing binaries: AR Lac, U CrB, S Equ, SU Boo, VV UMa, WW Gem, YY Eri, V0404 Lyr, HP Aur. All of them are well-known stellar systems and were observed during long period of time (80-150 years). To carry out our research, we took some general parameters (period, initial epoch, variability type etc.) of the systems from General Catalogue of Variable Stars (GCVS [1]). Masses of the binary systems were taken from previously published articles. All these parameters are collected in Table 1.

Unfortunately, only for three stellar systems (AR Lac, WW Gem and HP Aur) errors of components masses were provided. For all other stars we supposed that errors of masses for primary and secondary components are equal to 7\% of their masses. This value is equal to average relative error of stellar masses determination. In addition, there is a lot of information published before in other articles of different authors.

\begin{center}
Table 1. General information about investigated stars
\end{center}
\begin{center}
\begin{tabular}{|p{2.5cm}|p{2.5cm}|p{2.5cm}|p{2cm}|p{2cm}|p{1cm}|}
\hline
Stellar system & Initial epoch (JD-2400000) & Period (days) & $M_{1}, M_{\odot}$ & $M_{2}, M_{\odot}$ & Ref\\
\hline
AR Lac & 41593.7123 & 1.98319204 & 1.26 $\pm$ 0.02 & 1.12 $\pm$ 0.02 & [2]\\
\hline
U CrB & 16747.9718 & 3.45220133 & 4.8 & 1.4 & [3]\\
\hline
S Equ & 42596.74348 & 3.4360969 & 3.24 & 0.42 & [4]\\
\hline
SU Boo & 52500.895 & 1.561258 & 2.5 & 0.3 & [5]\\
\hline
VV UMa & 45815.3365 & 0.68738 & 1.93 & 0.44 & [6]\\
\hline
WW Gem & 25984.257 & 1.237811 & 4.39 $\pm$ 0.33 & 2.11 $\pm$ 0.16 & [6]\\ 
\hline
YY Eri & 41581.624 & 0.32149415 & 1.54 & 0.62 & [7]\\
\hline
V0404 Lyr & 35836.462 & 0.73094585 & 1.35 & 0.51 & [8]\\
\hline
HP Aur & 46353.236 & 1.4228191 & 0.9543 $\pm$ 0.0041 & 0.8094 $\pm$ 0.0036 & [9]\\
\hline
\end{tabular}
\end{center}
\subsection{AR Lac}
\begin{figure}[h]
\centering
\includegraphics[width=75mm]{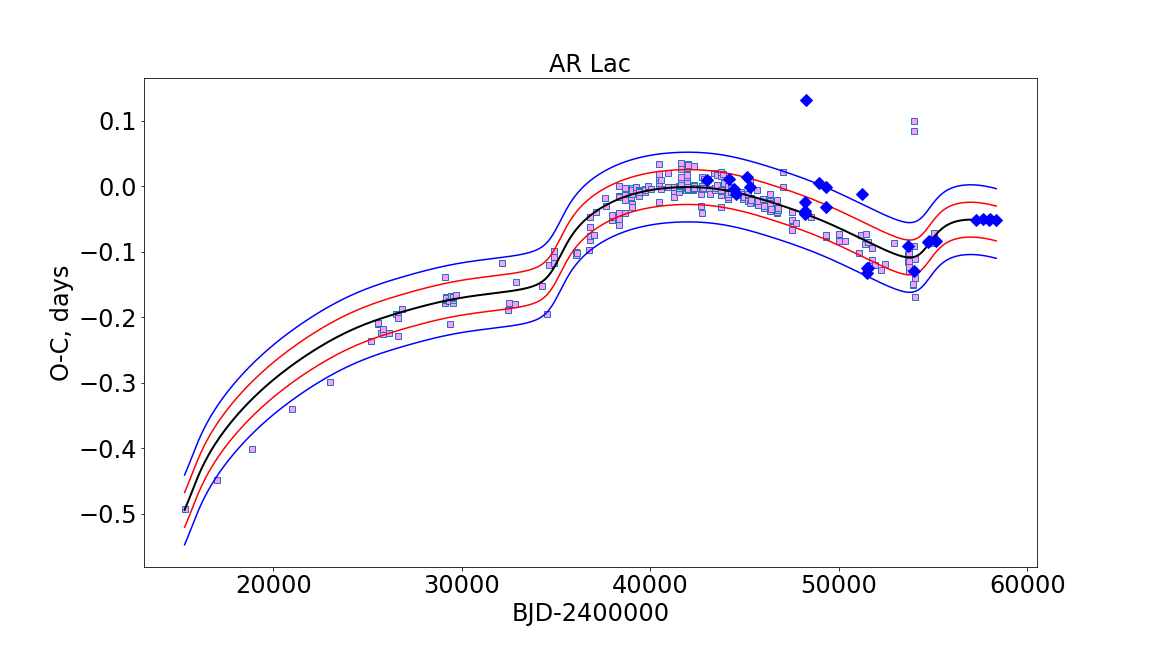}
\caption{O-C curve of AR Lac. For each star we plotted O-C curve and made approximation. Small dots are observations from database BRNO, large squares are ones that we determined from AAVSO observations. The lines show the approximation and the $\pm 1\sigma$ and $\pm 2\sigma$ ''error corridors'' , where $\sigma$ is a biased estimate of the r.m.s. deviation of a single point from the approximation.}
\end{figure}
There are more than 200 articles published before, thus there is analysis only of the most important of them. Third component was supposed in [10], [11], [12]. In all three articles mass of the third component was estimated. However, parameters of its orbit ware not computed.
Mass transfer was assumed in [11], [13], [14], [15] and [10]. In last four of these articles rate of the mass transfer was computed.

\subsection{U CrB}
\begin{figure}[h]
\centering
\includegraphics[width=75mm]{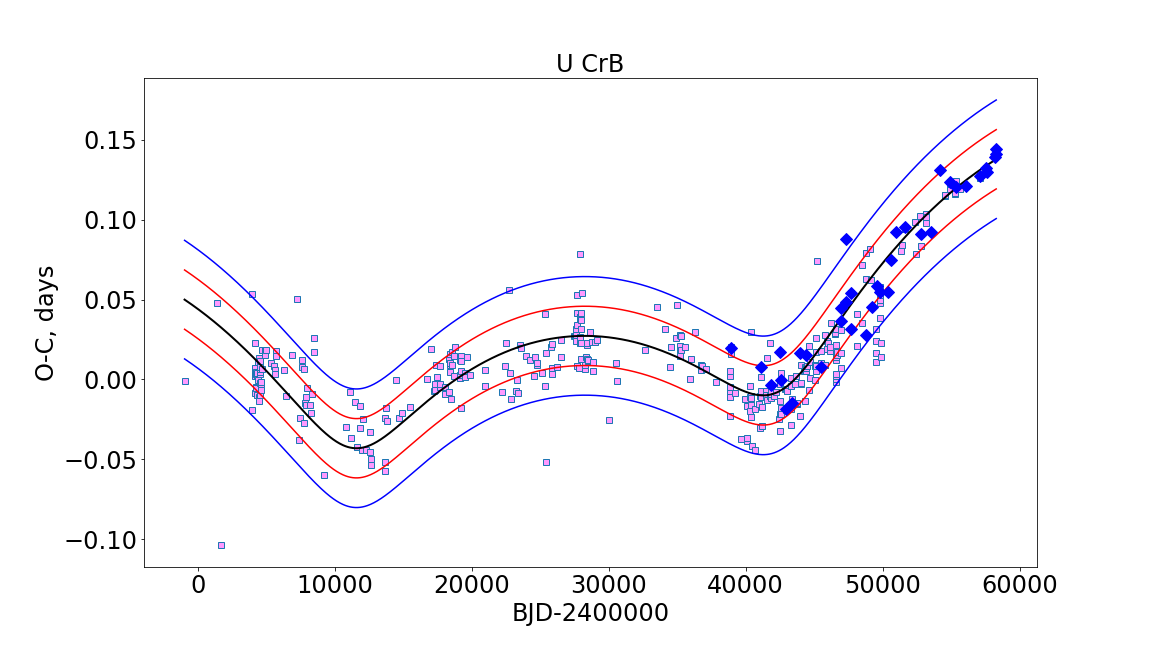}
\caption{O-C curve of U CrB}
\end{figure}
Mass transfer was supposed in [16], [17], [18], [19], [20], [21], [22], [23], [24] and [25].
Rate of  the mass transfer was computed in [16], [18], [21], [26] and [24].
Third component as the reason for cyclic period changes was supposed in [16], [26], [27], [28], [29], [31], [30].
Third component's mass was computed in [16], [29], [31], [30]. Orbital elements were estimated in [27] and [29]. Moreover, in [26] fourth component was supposed and its mass was computed.

\subsection{S Equ}
\begin{figure}[h]
\centering
\includegraphics[width=75mm]{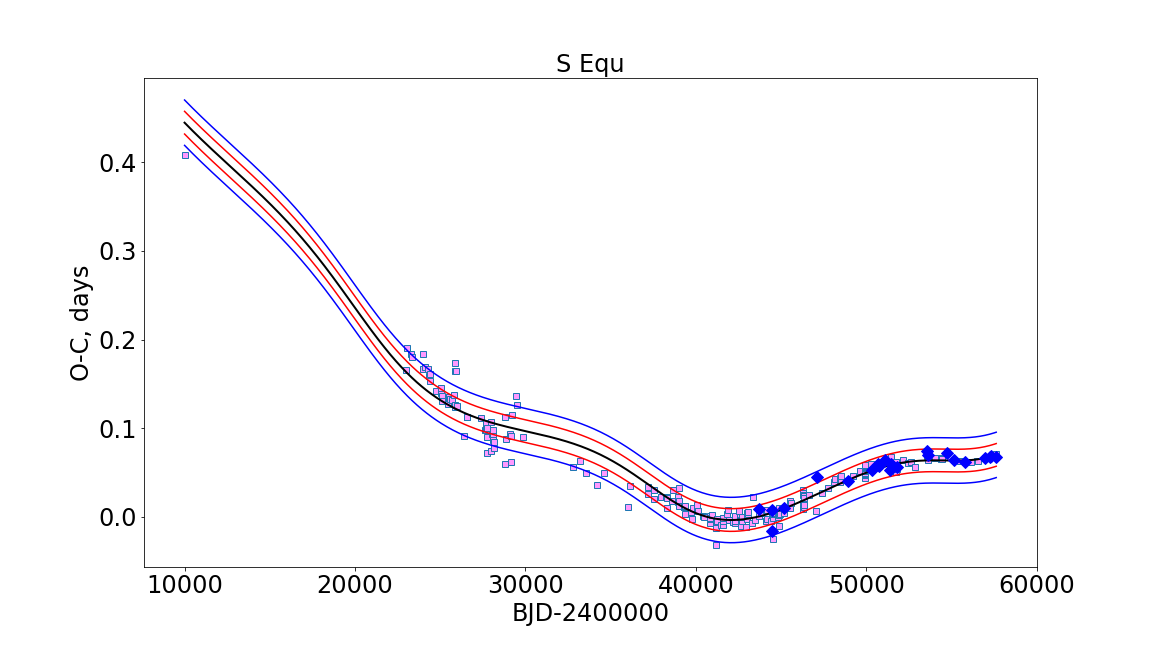}
\caption{O-C curve of S Equ}
\end{figure}
Mass transfer was supposed and its rate was determined in [31], [32], [33], [34], [35] and [36].
Third component was assumed and its mass was computed in [34], [35]. Moreover, in [34] parameters of the third component's orbit were calculated.

\subsection{SU Boo}
\begin{figure}[h]
\centering
\includegraphics[width=75mm]{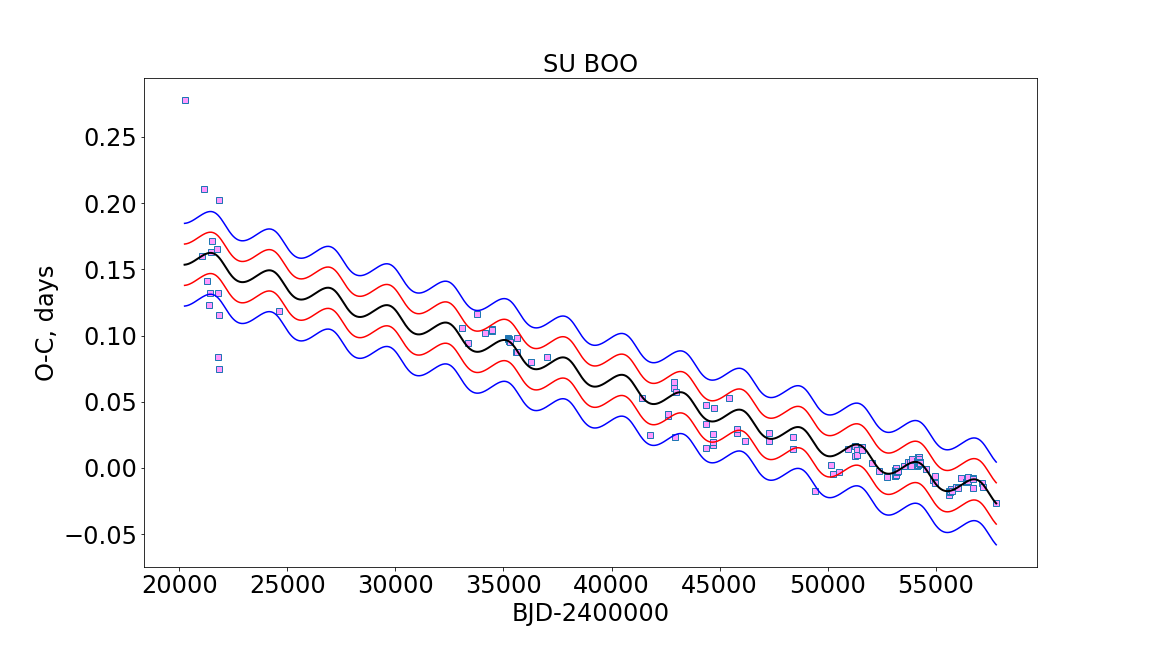}
\caption{O-C curve of SU Boo}
\end{figure}
In [37], complex analysis of the third component and mass transfer hypothesis was provided. Rate of the mass transfer was computed as well as the third component's mass and parameters of its orbit.

\subsection{VV UMa}
\begin{figure}[h]
\centering
\includegraphics[width=75mm]{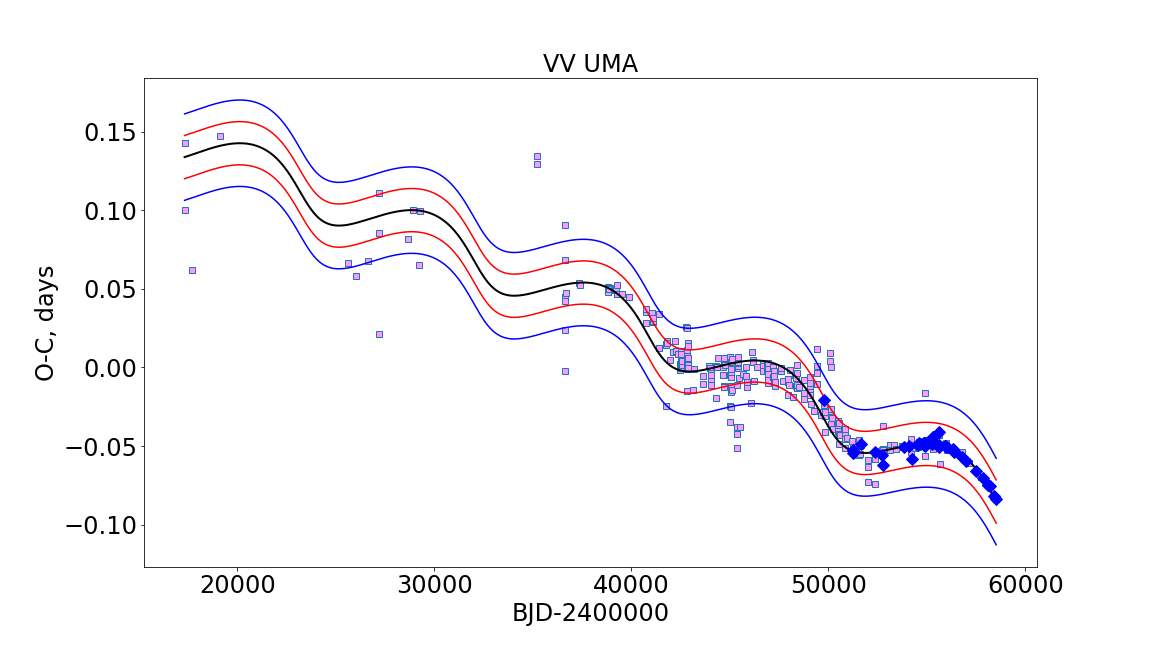}
\caption{O-C curve of VV UMa}
\end{figure}
In [37], hypothesis of the third component presence was considered; parameters of the orbit and mass were calculated.

\subsection{WW Gem}
\begin{figure}[h]
\centering
\includegraphics[width=75mm]{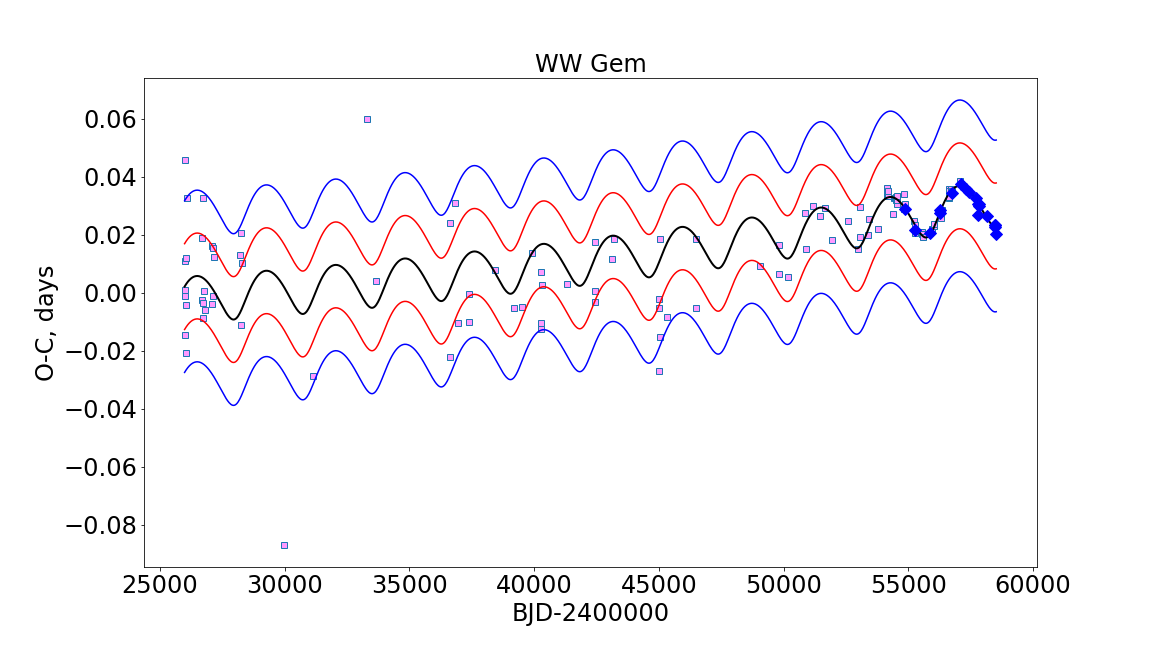}
\caption{O-C curve of WW Gem}
\end{figure}
Analysis of the third component and mass transfer was provided in [38]. Mass, orbital elements and mass transfer rate were determined.

\subsection{YY Eri}
\begin{figure}[h]
\centering
\includegraphics[width=75mm]{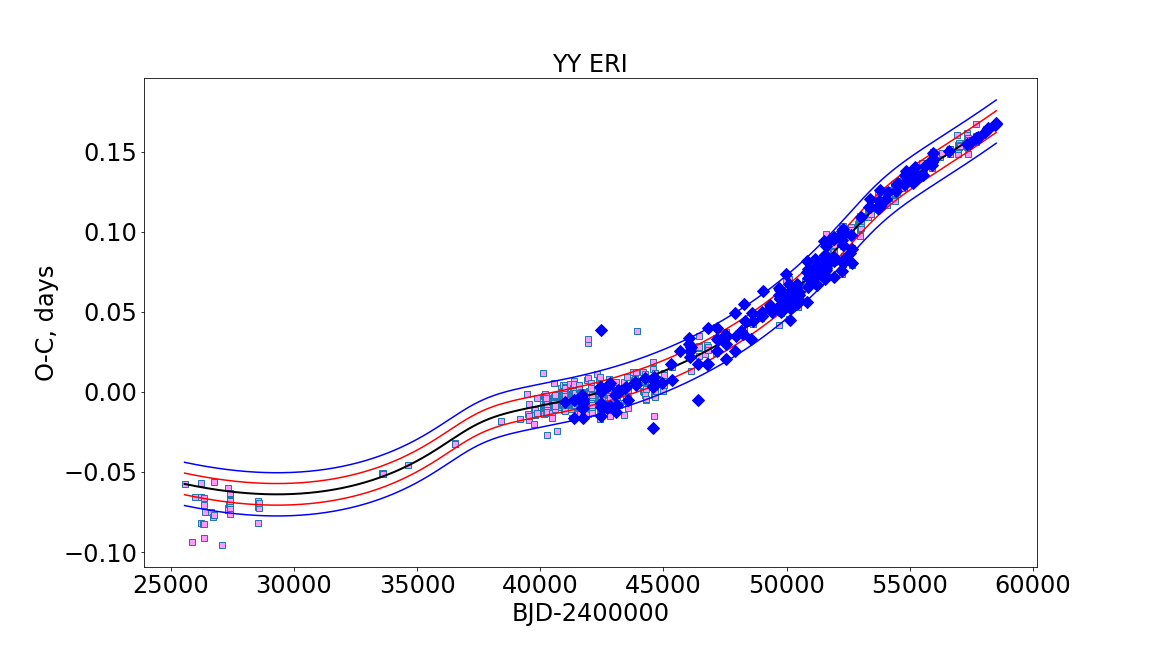}
\caption{O-C curve of YY Eri}
\end{figure}
In [39], third component was supposed and its mass was computed.

\subsection{V0404 Lyr}
\begin{figure}[h]
\centering
\includegraphics[width=75mm]{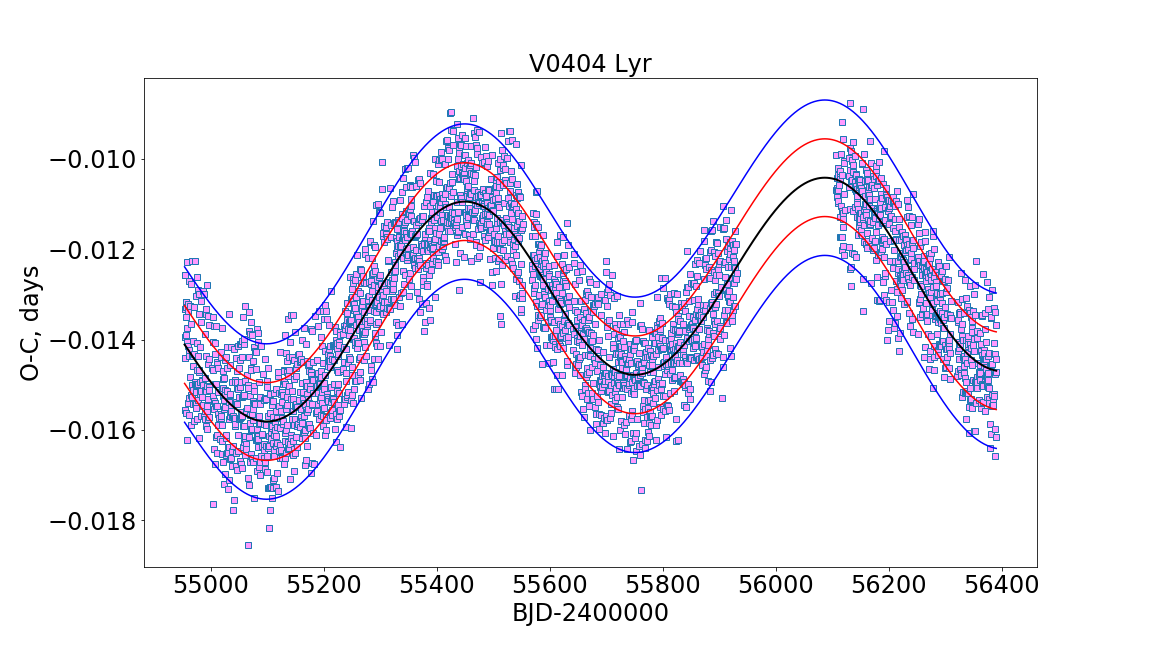}
\caption{O-C curve of V0404 Lyr}
\end{figure}
In [40] both third and fourth components were supposed and their masses obtained. Mass transfer rate was computed as well.

\subsection{HP Aur}
\begin{figure}[h]
\centering
\includegraphics[width=75mm]{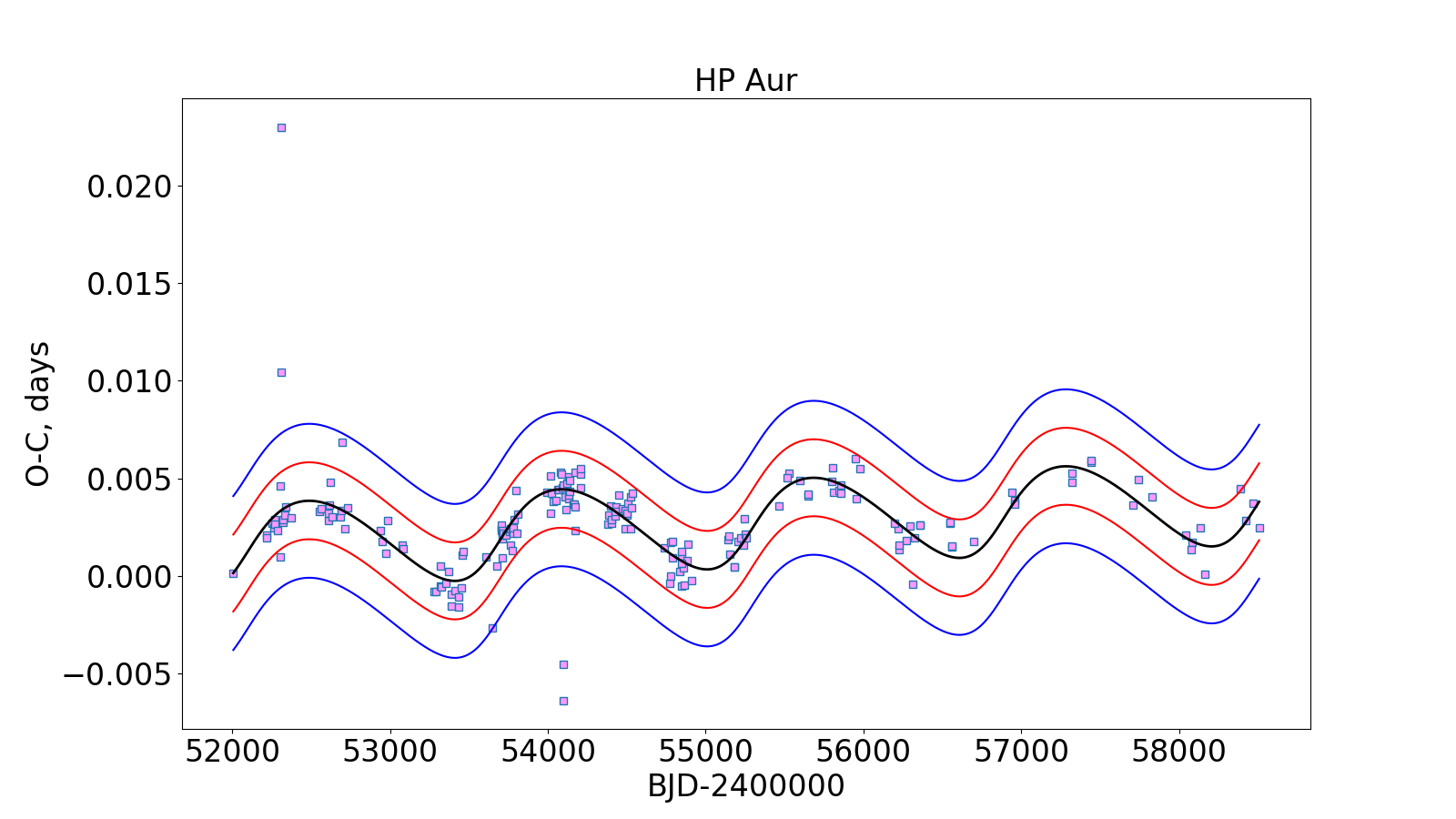}
\caption{O-C curve of HP Aur}
\end{figure}
No analysis of neither mass transfer nor of third component presence was provided in any article.
\section{Methods and algorithms}
One of the methods of third component's detection in eclipsing binary systems is O-C curve investigation. O-C curve is a dependence between observed moment of minimum minus calculated one and the time.

This method is one of the simplest when third component is neither visible nor taking part in the eclipses. It is appropriate for any possible mass of the third component if this mass is enough to make O-C oscillations larger then noise. In this case, we can suppose the third component presence without spectroscopic observations and even estimate some of its orbital parameters. On the other hand, third components usually have long orbital period. That is why several decades of regular observations are needed.

If O-C curve has sinusoid-like oscillations, we can suppose presence of the third component which makes the binary system to rotate around common barycenter. This motion causes delay (either positive or negative) of the minimal brightness moment. If we plot the dependence of the delay on time we get sinusoid-like O-C curve. After calculating amplitude and the period of such changes we can estimate minimal possible mass of the third component. Although, sometimes oscillations have clear asymmetry but stay periodic which might be caused by elliptical shape of the third component's orbit. Parameters of such which could be estimated by special algorithms.

In the case of parabolic shape of O-C or parabolic trend with superposition of cyclic period changes, we suppose presence of mass transfer between components of binary system. If O-C is linear it could be caused by the error in the period determination.

We estimated minimal possible mass of the third component using third Kepler's law and formula of the barycenter position. Then we supposed that orbit of the third component is perpendicular to the picture plane (thus we get the minimal possible value). After simplifications we got the international formula for minimal possible mass of the third component:
\begin{equation}
M_{3} = \frac{c \cdot \Delta t}{\sqrt[3]{G}} \cdot \left[\frac{2\pi}{T} \cdot (M_{1} + M_{2} + M_{3})\right]^{\frac{2}{3}}
\end{equation}
To compute third components mass using this formula several iterations are needed. Error of the minimal possible mass also was estimated:
\begin{equation}
\sigma M_{3} = \left(\frac{1}{M_{3}} - \frac{2}{3M}\right)^{-1} \cdot  \sqrt{ \left(\frac{\sigma \Delta t}{\Delta t}\right)^2 + \frac{4}{9} \left[\left(\frac{\sigma M_{1}}{M_{1}}\right)^2 + \left(\frac{\sigma M_{2}}{M_{2}}\right)^2 + \left(\frac{\sigma T}{T}\right)^2 \right]}
\end{equation}
The next stage is consideration of the mass transfer rate:
\begin{equation}
\dot{M} = \frac{1}{3} \frac{\dot{P}}{P} \frac{M_{1} M_{2}}{M_{1}-M_{2}}
\end{equation}
Error of the mass transfer rate:
\begin{equation}
\sigma \dot{M} = \dot{M} \sqrt{\left(\frac{\sigma P}{P}\right)^2 + \left(\frac{\sigma \dot{P}}{\dot{P}}\right)^2 + \left(\frac{M_{2}}{M_{1}} \frac{\sigma M_{1}}{M_{1}-M_{2}}\right)^2 + \left(\frac{M_{1}}{M_{2}} \frac{\sigma M_{2}}{M_{1}-M_{2}}\right)^2}
\end{equation}
Here:
\begin{itemize}
\item $M_{1}$, $M_{2}$, $M_{3}$ - masses of the components;
\item M - sum of the components masses;
\item P - period of variability;
\item $\dot{P}$ - rate of the period change (days per day);
\item $\dot{M}$ - rate of the mass transfer (solar masses per year);
\item c - speed of light in vacuum;
\item G - gravitational constant;
\item T, $\Delta t$ period and amplitude of the O-C oscillations;
\item $\sigma P$, $\sigma \dot{P}$, $\sigma M_{1}$, $\sigma M_{2}$, $\sigma \Delta t$, $\sigma T$ - errors of the period, rate of the period changes, masses of the binary system's components, amplitude and period of the O-C oscillations.
\end{itemize}
Rate of the period change could be obtained by approximation of the parabolic trend. The formula of parabola is well-known:
\begin{equation}
O-C = \alpha t^2 + \beta t + \gamma
\end{equation}
Here parameter
\begin{equation}
\alpha = \frac{\dot{P}}{2}
\end{equation}
is proportional to the rate of the period changes; $\beta$ is an error of the period in the moment t=0; $\gamma$ is the vertical shift (error of the initial epoch).
Finally, we considered the elliptic shape of the orbit. There are 6 main orbital elements which completely describe the third component's motion:
\begin{itemize}
\item semi-major axis (a);
\item eccentricity (e);
\item three angles of inclination
\begin{itemize}
\item argument of the pericenter ($\Omega$);
\item inclination (i);
\item longitude of pericenter ($\omega$);
\end{itemize}
\item orbital period (T);
\item moment of the pericenter transit ($t_{0}$).
\end{itemize}
\section{Data processing}
There were several stages in our research:
\begin{enumerate}
\item Collecting moments of minima from database BRNO [41];
\item Collecting amateur observations from AAVSO database [42];
\item Processing AAVSO observations using MAVKA software [43];
\begin{enumerate}
\item Splitting data onto separate minima;
\item Cutting out extraeclipsing part of the minima (if necessary);
\item Processing each individual minimum using symmetrical polynomial and moment of minima obtaining.
\end{enumerate}
\item Combination of AAVSO and BRNO minima;
\item Plotting O-C curve and estimation of the initial values of orbital parameters by eye;
\item Correction of the orbital parameters by modelling program;
\item Making plots of O-C curve and deviations for visual control;
\item Calculation of the third component's mass;
\item Calculation of the mass transfer rate;
\item Estimation of the errors.
\end{enumerate}
All results of calculations are provided in Appendix section in tables and pictures.
\newpage

\section{Discussion and conclusions}
\begin{center}
Table 2. Orbital elements, masses of the third components and mass transfer rates for all 9 stars
\end{center}
\begin{center}
\begin{tabular}{|p{1cm}|p{1cm}|p{1cm}|p{1cm}|p{1cm}|p{1cm}|p{1cm}|p{1cm}|p{1cm}|p{1cm}|p{1cm}|}
\hline
 & Units & AR Lac & U CrB & S Equ & SU Boo & VV UMa & WW Gem & YY Eri & V0404 Lyr & HP Aur\\
\hline
$\alpha$ & $10^{-12} \newline days^{-1}$ & -425 $\pm$ 8 & 77 $\pm$ 8 & 382 $\pm$ 5 & & & & 212 $\pm$ 2 & & \\
\hline
$\beta$ & $10^{-7}$ & 391 $\pm$ 7 & -299 $\pm$ 2 & -336 $\pm$ 4 & -46.9 $\pm$ 0.3 & -58.6 $\pm$ 0.2 & 12.0 $\pm$ 0.7 & -109.1 $\pm$ 1.4 & 11.17 $\pm$ 0.05 & 4.26 $\pm$ 0.1\\
\hline
$\gamma$ & $10^{-3} \newline days$ & -957 $\pm$ 15 & 3 $\pm$ 3 & 751 $\pm$ 9 & 252 $\pm$ 2 & 261 $\pm$ 1 & -41 $\pm$ 4 & 85 $\pm$ 3 & -75.0 $\pm$ 0.3 & -21.2 $\pm$ 0.5\\
\hline
$a \sin{i}$ & $10^6 \newline km$ & 1420 $\pm$ 30 & 914 $\pm$ 15 & 548 $\pm$ 12 & 216 $\pm$ 7 & 419 $\pm$ 9 & 207 $\pm$ 7 & 271 $\pm$ 2 & 56.75 $\pm$ 0.06 & 56.9 $\pm$ 0.6\\
\hline
e & $10^{-3}$ & 470 $\pm$ 27 & 389 $\pm$ 30 & 137 $\pm$ 41 & 264 $\pm$ 47 & 325 $\pm$ 30 & 262 $\pm$ 64 & 416 $\pm$ 12 & 79 $\pm$ 2 & 153 $\pm$ 20\\
\hline
$\omega$ & rad & -6.63 $\pm$ 0.04 & 4.79 $\pm$ 0.06 & 3.27 $\pm$ 0.23 & 0.35 $\pm$ 0.23 & 3.51 $\pm$ 0.07 & -1.08 $\pm$ 0.26 & 0.84 $\pm$ 0.03 & 2.78 $\pm$ 0.03 & -7.38 $\pm$ 0.13\\
\hline
$t_{0}$ & JD-2400000 & -21390 $\pm$ 610 & -19430 $\pm$ 590 & 3210 
$\pm$ 790 & 4070 $\pm$ 200 & 6120 
$\pm$ 220 & 3010 $\pm$ 300 & 4610 $\pm$ 190 & 4980 $\pm$ 20 & 10560 $\pm$ 80\\
\hline
T & days & 19000 $\pm$ 140 & 30930 $\pm$ 280 & 17380 $\pm$ 180 & 3088 $\pm$ 9 & 8750 $\pm$ 30 &  2783 $\pm$ 15 & 16230 $\pm$ 60 & 640 $\pm$ 0.2 & 1590 $\pm$ 3\\
\hline
$\dot{M}$ & $10^{-9} \newline \frac{M_{\odot}}{year}$ & 133 $\pm$ 27 & 1.55 $\pm$ 0.16 & 1.90 $\pm$ 0.15 & - & - & - & 259 $\pm$ 31 & - & -\\
\hline
$M_{3}$ & $M_{\odot}$ & 1.746 $\pm$ 0.067 & 1.204 $\pm$ 0.089 & 0.750 $\pm$ 0.057 & 4.795 $\pm$ 0.523 & 0.820 $\pm$ 0.069 & 0.714 $\pm$ 0.056 & 0.530 $\pm$ 0.040 & 0.260 $\pm$ 0.130 & 0.457 $\pm$ 0.033\\
\hline
\end{tabular}
\end{center}

\begin{figure}[h]
\centering
\includegraphics[width=75mm]{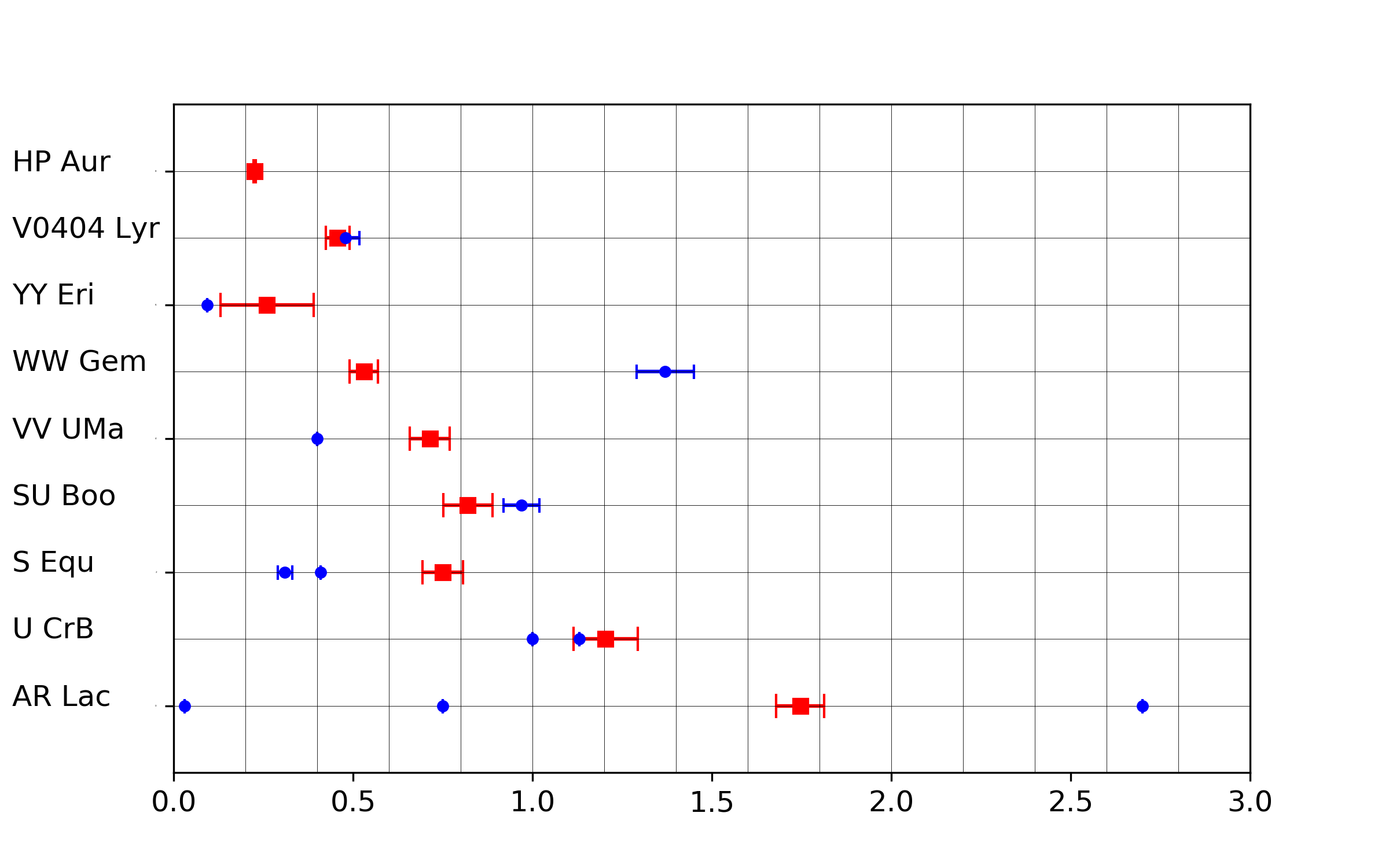}
\caption{Computed masses of the third components (squares) in comparison with previous results (dots)}
\end{figure}

Our results generally are not in agreement with results of other authors. Only for V0404 Lyr our estimation is more or less the same than in previos research. For many stars our values are much higher - probably, they are overestimated. However, for WW Gem and SU Boo masses are smaller, so we could underestimate them. Unfortunately, only several authors provided errors of their calculations that makes the comparison process really difficult. Anyway, the order of value remains the same for all 3rd components' masses and we used much larger data sets. Thus, probably, our data is more accurate, but we need further observations to prove that thesis.
\section{Acknowledgments}
We sincerely thank the AAVSO (Kavka, 2018) and BRNO associations of variable stars observers for their work that has made this research possible. In addition we are grateful to Andrych et al. who provided us their software MAVKA for obtaining moments of minima [43], [44], [45]. This research was done as the part of the projects Inter-Longitude Astronomy [49], UkrVO [50] and AstroInformatics [51], [52] as well as previous research [53], [54], [55]. Special acknowledgments to Leonid Shakun for constructive discussion about algorithm of the Python program.
\section{References}
\begin{enumerate}
\item N.N. Samus, E.V. Kazarovets, O.V. Durlevich, N.N. Kireeva, E.N. Pastukhova, Astronomy Reports, 61, 80-88 (2017), http://www.sai.msu.su/gcvs/; DOI: 10.1134/S1063772917010085
\item M. Zboril, J.M. Oliveira, S. Messina, G. Djurasevic, P.J. Amado, Contrib. Astron. Obs. Skalnate Pleso 35, 23-34 (2005);
\item E. Raumer, Monthly Notices of the Royal Astronomical Society, 427, 1702-1712 (2012);
\item J.P. De Greve, N. Mennekens, W. Van Rensbergen, L. Yungelson, ASP Conference Series, 404 (2009);
\item F. Mardirossian, M. Mezzetti, G. Giuricin, Astronomy and Astrophysics Supplement Series, 40, 57-66 (1980);
\item C. Lazaro, M. J. Arevalo P A. Claret P E. Rodriguez P, I. Olivares Monthly Notices of the Royal Astronomical Society, 325, 617-630 (2001);
\item Y.-G. Yang, Y. Yang, H.-F. Dai, X.-G. Yin, The Astronomical Journal, 148:90 (5pp) (2014);
\item R. Nesci, C. Maceroni, L. Milano, G. Russo, Astronomy and Astrophysics, 159, 142-146 (1986);
\item Jae Woo Lee, The Astronomical Journal, 148:37 (11pp) (2014);
\item D. S. Hall, J. M. Kreiner, Acta Astronomica, 30, 3, 387-451 (1980);
\item Chun-Hwey Kim, The Astronomical Journal, 102, 5, 1784-1789 (1991);
\item L. Jetsu et al, Astronomy and Astrophysics, 326, 698-708 (1997);
\item J. D. Needham, J. P. Phillips, M. J. Selby, C. Sanchez-Margo, Astronomy and Astrophysics, 83, 370-374 (1980);
\item R.K. Srivastava, Acta Astronomica, 34, 2, 291-302 (1984);
\item Ye Lu, Fu-Yuan Xiang, Xiao-Min Shi, Publ. Astron. Soc. Japan, 64, 4, 84-1 - 84-8 (2012);
\item A.I. Khaliullina, Astronomy Reports, 2018, 62, 4, 264-272 (2018);
\item M.T. Richards and A.S. Cocking, Proceedings IAU Symposium 290, 301-302 (2012);
\item E. Raymer, Mon. Not. R. Astron. Soc. 427, 1702-1712 (2012);
\item M.T. Richards, Proceedings IAU Symposium 282, 167-172 (2012);
\item M.I. Agafonov, O.I. Sharova, M.T. Richards, The Astrophysical Journal, 690, 1730-1744 (2009);
\item S. K. Yerli et al, Mon. Not. R. Astron. Soc. 342,1349-1360 (2003);
\item G.J. Peters, R.S. Polidan, Bulletin of the American Astronomical Society, 29, 835 (1997);
\item G.E. Albright and M.T. Richards, Astrophysics and Space Science, 224, 415-416 (1995);
\item M.T. Richards, G.E. Albright, L.M. Bowles, The Astrophysical Journal, 438, L103-L106 (1995);
\item J. Tomkin, D.L. Lambert, M.Lemke, Mon. Not. R. Astron. Soc., 265, 581-587 (1993);
\item T. Borkovits, T. Hegedus, Odessa Astronomical Publications, 7, 2, 126 (1994);
\item P. Mayer, M. Wolf, J. Tremko, P.G. Niarchos, Publishing House of the Czechoslovak Academy of Science, 42, 4 (1991);
\item J. R. W. Heintze, AAVSO, 19, 23-27 (1990);
\item R.H. van Gent, Astronomy and Astrophysics Supplement Series, 77, 471-485 (1989);
\item H. Forbes-Conde and T. Hertczeg, Astronomy and Astrophysics Supplement Series, 12, 1-78 (1973);
\item G.A. Bakos and J.Tremko, The Royal Astronomical Society of Canada, 75, 3, 124-131 (1981);
\item N. Mennekens, J.-P. De Greve, W. Van Rensbergen, L. R. Yungelson, Astronomy and Astrophysics, 486, 919-921 (2008);
\item F. Soydugan et al., Mon. Not. R. Astron. Soc., 379, 1533-1545 (2007);
\item F. Soydugan, O. Demircan, E. Soydugan, C.Ibanoglu, The Astronomical Journal, 126, 393-397 (2003);
\item S.B. Qian and L.Y. Zhu, The Astrophysical Journal Supplement Series, 142, 139-143 (2002);
\item M.T. Richards and G.E. Albright, The Astrophysical Journal Supplement Series, 123, 537-626 (1999);
\item P. Zasche, M. Wolf, R. Uhlar, H. Kucakova, The Astronomical Journal, 147, 130 (9pp) (2014);
\item V. Simon, Astronomy and Astrophysics, 311, 915-918 (1996);
\item L.F. Snyder, Society for Astronomical Sciences, 32nd Annual Symposium on Telescope Science..179-184 (2013), ADS: 2013SASS...32..179S;
\item Jae Woo Lee, Seung-Lee Kim, Kyeongsoo Hong, Chung-Uk Lee, Jae-Rim Koo, arXiv:1405.5658v1 [astro-ph.SR] (2014);
\item Brno Regional Network of Observers, http://var2.astro.cz/EN/;
\item American Association of Variable Stars Observers, https://www.aavso.org/;
\item K.D. Andrych, I.L. Andronov, Open European Journal On Variable Stars, 197, 65-70 (2019);
\item K.D. Andrych, I.L. Andronov, L.L. Chinarova, Odessa Astronomical Publications, 30, 57-62 (2017);
\item K.D. Andrych, I.L. Andronov, L.L. Chinarova, V.I. Marsakova, Odessa Astronomical Publications, 28, 158-160 (2015);
\item J.B. Tatum, Physics topics. Celestial Mechanics: http://astrowww.phys.uvic.ca/\tt{~}tatum/celmechs/;
\item Scipy library: https://www.scipy.org/;
\item Python: http://www.python.org/;
\item I.L. Andronov, K.D. Andrych, K.A. Antoniuk, et al., ASP Conf. Ser., 511, 43-50 (2017), ADS: 2017ASPC..511...43A
\item I.B.Vavilova, Ya.S. Yatskiv, L.K. Pakuliak, et al., IAU Symp., 325, 361-366 (2017), DOI: 10.1017/S1743921317001661
\item I.B. Vavilova, L.K. Pakulyak, A.A. Shlyapnikov, et al., Kinematics and Physics of Celestial Bodies, 28, 2, 85-102 (2012),  DOI: 10.3103/S0884591312020067
\item Вавилова И.Б., Пакуляк Л.А., Шляпников АА , Процюк Ю.И. и др., 2012, Кинематика и физика небесных тел (in Russian), 28, 2, 85
\item D.E. Tvardovskyi, V.I. Marsakova, I.L. Andronov, Odessa Astronomical Publications, 30, 135-139 (2017);
\item D.E. Tvardovskyi, V.I. Marsakova, I.L. Andronov, L.S. Shakun, Odessa Astronomical Publications, 31, 103-109 (2018);
\item D.E. Tvardovskyi, Pre-print: arXiv:1911.12415 (Advances in Astronomy and Space Physics, in press.)
\end{enumerate}
\end{document}